\documentclass[preprint,12pt, a4paper]{elsarticle}

\usepackage{xspace}
\usepackage{amsmath}
\usepackage{amssymb}
\usepackage{hyperref}
\usepackage{onimage}
\usepackage{subcaption}
\usepackage{listings}
\usepackage{tabularx}
\journal{SoftwareX}

\DeclareGraphicsExtensions{.png}

\newcommand{\resforpy}{\texttt{resolvent4py}\xspace}
\newcommand{\petscforpy}{\texttt{petsc4py}\xspace}
\newcommand{\slepcforpy}{\texttt{slepc4py}\xspace}
\newcommand{\mpiforpy}{\texttt{mpi4py}\xspace}

\newcommand{\revone}[1]{\textcolor{black}{#1}}
\newcommand{\revtwo}[1]{\textcolor{black}{#1}}

\newtheorem{remark}{Remark}

\begin{document}
\renewcommand{\labelenumii}{\arabic{enumi}.\arabic{enumii}}

\begin{frontmatter}



\title{Resolvent4py: a parallel Python package for analysis, model reduction and control of large-scale linear systems}


\author[uiuc]{Alberto Padovan}
\author[caltech]{Vishal Anantharaman}
\author[princeton]{Clarence W. Rowley}
\author[uiuc]{Blaine Vollmer}
\author[caltech]{Tim Colonius}
\author[uiuc]{Daniel J. Bodony}
\address[uiuc]{Department of Aerospace Engineering, University of Illinois Urbana-Champaign, 104 S. Wright St., Urbana, IL 61802}
\address[caltech]{Department of Mechanical and Civil Engineering, California Institute of Technology, 1200 E. California Boulevard, Pasadena, CA 91125}
\address[princeton]{Department of Mechanical and Aerospace Engineering, Princeton University, Olden St., Princeton, NJ 08540}

\begin{abstract}
In this paper, we present \resforpy, a parallel Python package for the analysis, model reduction and control of large-scale linear systems with millions or billions of degrees of freedom.
This package provides the user with a friendly Python-like experience (akin to that of well-established libraries such as \texttt{numpy} and \texttt{scipy}), while enabling MPI-based parallelism through \mpiforpy, \petscforpy and \slepcforpy. 
In turn, this allows for the development of streamlined and efficient Python code that can be used to solve several problems in fluid mechanics, solid mechanics, graph theory, molecular dynamics and several other fields.
\end{abstract}


\end{frontmatter}


\section*{Metadata}


\begin{table}[!h]
\caption{Code metadata}
\label{codeMetadata} 
\begin{tabularx}{\textwidth}{|l|X|X|}
\hline
\textbf{Nr.} & \textbf{Code metadata description} & \textbf{Metadata} \\
\hline
C1 & Current code version & v1.0.1 \\
\hline
C2 & Permanent link to code/repository used for this code version & \url{https://github.com/albertopadovan/resolvent4py} \\
\hline
C3 & Permanent link to Reproducible Capsule & -- \\
\hline
C4 & Legal Code License & MIT \\
\hline
C5 & Code versioning system used & Git (GitHub) \\
\hline
C6 & Software code languages, tools, and services used & Python, Petsc4py, Slepc4py, Mpi4py \\
\hline
C7 & Compilation requirements, operating environments \& dependencies & See README file in package distribution \\
\hline
C8 & If available Link to developer documentation/manual & \url{https://albertopadovan.github.io/resolvent4py/} \\
\hline
C9 & Support email for questions & \href{mailto:alberto.padovan.94@gmail.com}{alberto.padovan.94@gmail.com} \\
\hline
\end{tabularx}
\end{table}

\section{Motivation and significance}

The development of this package is motivated by recent (and not-so-recent) advances in the field of fluid mechanics, where linear analysis has been shown to provide great insights into flow physics.
For example, linear systems theory can be used to study the stability of fluid flows around equilibria, develop feedback control strategies to modify the flow behavior and compute reduced-order models to accelerate physical simulations.
\revtwo{(We refer the reader to the foundational work of \cite{Schmid2001stability} and to the review papers in \cite{Rowley2017model, jovanovix2021} for a more in-depth overview of the importance and impact of linear systems theory in the field of fluid mechanics.)}
For systems of moderate dimension (e.g., fewer than 10,000 states), all these tasks can be performed straightforwardly with a few lines of Python code thanks to scientific computing libraries like \texttt{scipy}.
For larger systems, the need for \revtwo{distributed-memory} parallelism significantly increases the implementation complexity of these algorithms, and it often requires the development of application-specific software written in compiled languages such as C, C++ and Fortran.
\revtwo{In fact, until the recent development of Python libraries like \texttt{mpi4py} and \texttt{petsc4py}, compiled languages represented the only way to write MPI-based\footnote{\revtwo{MPI stands for Message Passing Interface, and it refers to the communication protocol used to exchange data across different processors in distributed-memory systems.}} code.}
The objective of this package is to \revtwo{address these issues and} provide a user-friendly pythonic environment similar to that of \texttt{scipy} and \texttt{numpy}, while exploiting the large-scale parallelism offered by the \texttt{mpi4py}, \texttt{petsc4py} and \texttt{slepc4py} libraries.
In turn, this enables the development of user-friendly, streamlined and efficient code to perform day-to-day engineering tasks and physical analyses of large-scale systems.

\begin{remark}
    Although the development of this package is motivated by problems in the field of fluid mechanics, \resforpy is application agnostic, and its functionalities can be leveraged for any linear system of equations regardless of their origin.
\end{remark}


\section{Software description}

\resforpy is a Python library for the analysis, model reduction and control of large-scale linear systems.
It relies on \mpiforpy \cite{Dalcin2011} for distributed-memory parallelism, and it leverages the data structures and functionalities provided by the \petscforpy \cite{Balay2025} and \slepcforpy \cite{slepc2005} libraries.

\subsection{Software architecture}

At the core of this package is an abstract class, called \texttt{LinearOperator}, which serves as a blueprint for user-defined child classes that can be used to define a linear operator $L$.
All child classes must implement at least three methods: \texttt{apply()}, which defines the action of $L$ on a vector, \texttt{apply\_mat()}, which defines the action of $L$ on a matrix and \texttt{destroy()} to free the memory.
The implementation of the second method may seem redundant to the reader who is familiar with native Python, where the~\texttt{dot()} method can be used to apply the action of a linear operator to vectors and matrices alike. 
However, the need for this method is dictated by the fact that matrix-vector products and matrix-matrix products are performed with different functions and data structures in \petscforpy.

If desired, additional methods can be implemented.
For example, we often define the action of the hermitian transpose of a linear operator on vectors and matrices (\texttt{apply\_hermitian\_transpose()} and \texttt{apply\_hermitian\_trans\\pose\_mat()}, respectively), the action of the inverse $L^{-1}$ on vectors, \texttt{solve()}, \revtwo{etc.}
Below is a list of concrete subclasses of \texttt{LinearOperator} that are provided by \texttt{resolvent4py}:
\begin{enumerate}
    \item \texttt{MatrixLinearOperator}: the linear operator $L$ is defined by a sparse, MPI-distributed PETSc matrix.
    \item \texttt{LowRankLinearOperator}: the linear operator $L$ is defined as the product of low-rank factors,
    \begin{equation}
        L = U\Sigma V^*,
    \end{equation}
    where $U$ and $V$ are ``tall and skinny'' matrices \revtwo{(i.e., matrices with much fewer columns than rows)} stored as MPI-distributed SLEPc \texttt{BV} structures\footnote{The SLEPc \texttt{BV} structure (where BV stands for ``Basis Vector'') holds a tall and skinny dense matrix whose rows are distributed across different MPI processors.
    The columns, however, are not distributed, and this allows for significantly reduced parallel overhead when the number of MPI processors is much larger than the number of columns, as is usually the case in practice.
    }, while $\Sigma$ is a sequential \revtwo{(i.e., non-distributed)} two-dimensional \texttt{numpy} array of \revtwo{appropriate} size.
    \item \texttt{LowRankUpdatedLinearOperator}: the linear operator $L$ is defined as $L = A + M$, where $A$ is itself a \resforpy linear operator and~$M$ is a low-rank linear operator, as described in the previous bullet point. 
    An operator of this form arises in several applications \revtwo{(e.g., in linear feedback control theory \cite{natarajan2016actuator})} and, if $A$ and $L$ are full-rank, then the inverse $L^{-1}$ can be computed using the Woodbury matrix inversion lemma \cite{woodbury1950},
    \begin{equation}
        L^{-1} = A^{-1} - A^{-1}U\Sigma\left(I + V^*A^{-1}U\Sigma\right)^{-1}V^*A^{-1}.
    \end{equation}
    \item \texttt{ProjectionLinearOperator}: the linear operator $L$ is defined either as $L = \Phi \left(\Psi^*\Phi\right)^{-1}\Psi^*$ or $L = I - \Phi \left(\Psi^*\Phi\right)^{-1}\Psi^*$, where $\Phi$ and $\Psi$ are tall and skinny matrices of size $N \times r$ stored as SLEPc \texttt{BV}s.
    In both cases,~$L$ defines a projection (i.e., $L^2 = L$). 
    \item \texttt{ProductLinearOperator}: the linear operator $L$ is defined as the product of other \resforpy linear operators $L_i$ of conformal dimensions,
    \begin{equation}
        L = L_m L_{m-1}\ldots L_2 L_1.
    \end{equation}
\end{enumerate}

\subsection{Software functionalities}

Once a linear operator is properly defined, \resforpy currently allows for several types of analyses. 

\begin{enumerate}
    \item Eigendecomposition: the user can compute the left and right \revtwo{eigendecomposition} of the linear operator $L$ using the Arnoldi algorithm \cite{arnoldi1951}. The shift-and-invert technique \revtwo{\cite{saad2011numerical}} can be used to compute the eigenvalues that are closest to a target value $s\in\mathbb{C}$, where $\mathbb{C}$ is the set of complex numbers.
    \item Singular value decomposition (SVD): the user can compute an SVD of the linear operator using randomized linear algebra \cite{rsvd, ribeiro2020}.
    This is useful for resolvent analysis \cite{Schmid2001stability,Jovanovic2005componentwise, McKeon2010critical} and harmonic resolvent analysis \cite{Padovan2020analysis, padovan2022prf, wu2022prf, Islam_Sun_2024}.
    For the specific case of resolvent analysis, the package also offers the possibility of computing the SVD using time-stepping techniques \cite{Martini2021, farghadan2025}.
    \item Linear time-invariant balanced truncation: given the linear operator~$L$, and appropriately-defined input and output matrices $B$ and $C$, the user may compute and balance the associated controllability and observability Gramians for model reduction purposes \cite{Moore1981principal,Dullerud2000robust,Rowley2005model,Baur2014model}. Specifically, we implement the algorithm outlined in \cite{dergham2011}.
    \revone{We remark that there exist additional model reduction formulations for linear system, (see, e.g., \cite{Gugercin2008H2, Flagg2013, Benner2015survey, benner2005dimension} for $\mathcal{H}_2$ and $\mathcal{H}_\infty$ model reduction and variations thereof), but these are not yet implemented in the package.}
\end{enumerate}

Additionally, \resforpy ships with several functions---available under the \texttt{resolvent4py/utils} directory and directly accessible to the user via the \texttt{resolvent4py} namespace---that further facilitate the use of our package and allow for streamlined application-specific code development.
These include support for:
\begin{enumerate}
    \item parallel I/O through \texttt{petsc4py},
    \item MPI-based communications using \texttt{mpi4py},
    \item manipulation of PETSc matrices/vectors and SLEPc BVs.
\end{enumerate}
All these features are thoroughly documented using \texttt{sphinx}, and demonstrated with several examples that ship with the package.

\subsection{Sample code snippet}
\label{subsec: code_snippet}

In this subsection, we show a code snippet similar to the one used in section \ref{subsec: cone} to perform resolvent analysis on the hypersonic flow over a cone.
The first thing performed by this piece of code is to read in the sparse complex-valued matrix $A$ of size $N\times N$.
This matrix is stored on disk in sparse COO format by means of PETSc-compatible binary files \texttt{"rows.dat"}, \texttt{"columns.dat"} and \texttt{"values.dat"}.
After assembling $A$ as a sparse PETSc matrix, we create a corresponding \texttt{MatrixLinearOperator} object and we perform resolvent analysis using the RSVD-$\Delta t$ algorithm presented in \cite{Martini2021, farghadan2025}.
The desired singular values and vectors of the resolvent operator are then saved to file using the parallel I/O routines available under the \resforpy namespace.
Running this piece of code with, e.g., $20$ processors can be done with the command \texttt{mpiexec -n 20 python code\_snippet.py}.

\begin{lstlisting}[language=Python, caption=Resolvent analysis via time stepping]
import resolvent4py (*@\textcolor{blue}{as}@*) res4py

# Define problem size and read jacobian from file
N = 121404 # Number of global rows (and columns) of A
Nloc = res4py.compute_local_size(N) # Number of local rows
file_names = ["rows.dat", "columns.dat", "values.dat"]
A = res4py.read_coo_matrix(file_names, ((Nloc, N), (Nloc, N)))

# Create resolvent4py linear operator
L = res4py.linear_operators.MatrixLinearOperator(A)

# Define the inputs to the rsvd-dt routine
dt = 1e-5       # Time step size
omega = 1.0     # Fundamental frequency (sets period T = 2pi/omega)
n_omegas = 5    # Number of harmonics of the fundamental to resolve
n_periods = 100 # Number of periods to integrate through
n_loop = 1      # Number of power iterations
n_rand = 30     # Number of random vectors
n_svals = 3     # Number of singular values/vectors to resolve
tol = 1e-3      # Tolerance to declare that transients have decayed
verbose = 1     # Verbosity level
Ulst, Slst, Vlst = res4py.linalg.resolvent_analysis_rsvd_dt(L, dt, omega, n_omegas, n_periods, n_loops, n_rand, n_svals, tol, verbose)

# Save to file
np.save("S.npy", Slst[1])
res4py.write_to_file("U.dat", Ulst[1])
res4py.write_to_file("V.dat", Vlst[1])

\end{lstlisting}

\revtwo{
\begin{remark}
    We would like to stress how this snippet exposes the application-agnostic nature of our package.
    The governing equations (e.g., the linearized fluid dynamics equations, the linearized molecular dynamics equations, or any other linear equation from other branches of physics) are embedded in the sparse matrix $A$ that is passed as an input to the script.
    Once $A$ is loaded from file, \texttt{resolvent4py} performs the desired analysis without additional knowledge of the underlying physics.
\end{remark}}

\section{Illustrative examples}

\subsection{Hypersonic flow over a compression ramp}
\label{subsec: ramp}

\begin{figure}
\centering
\includegraphics[trim=15 20 20 130, clip, width=0.7\linewidth]{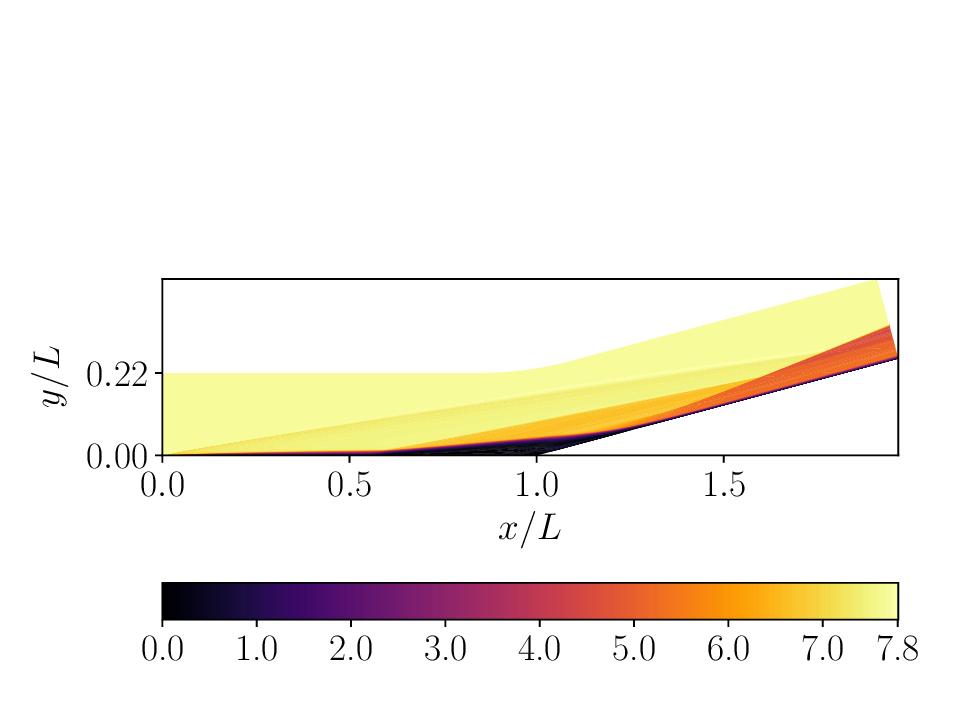}
\caption{Mach number contours of two-dimensional steady state solution for the Mach--~$7.7$ flow over a $15\deg$ compression ramp.}
\label{fig:ramp_mach_contours}
\end{figure}

\begin{figure}
\centering
\includegraphics[trim=10 10 10 10, clip, width=0.7\linewidth]{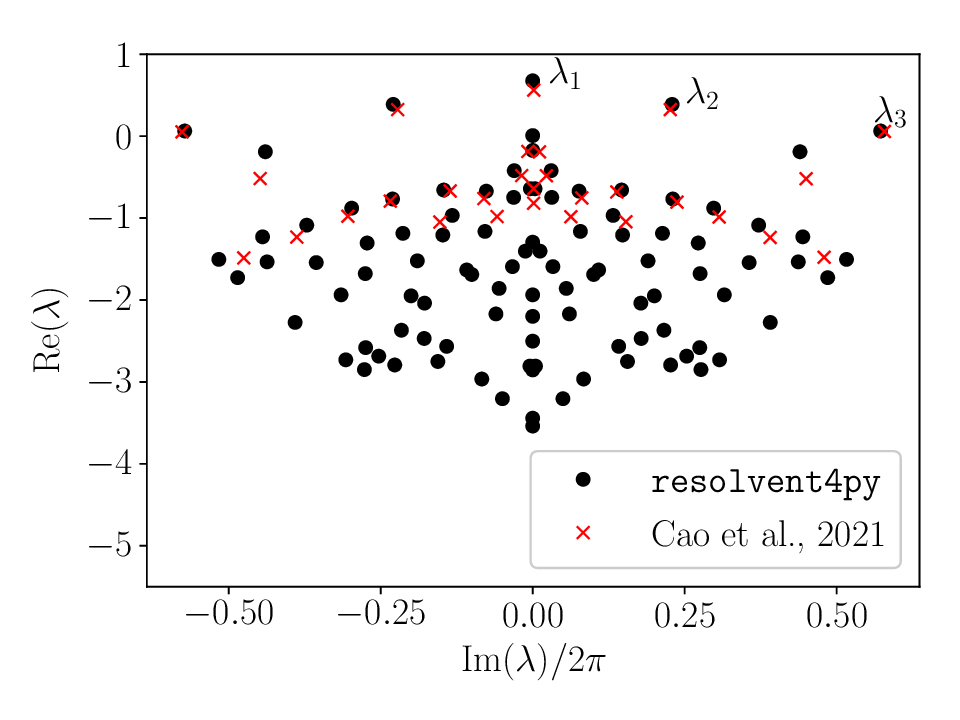}
\caption{Eigenvalues of the linearized Navier-Stokes equations at non-dimensional spanwise wavenumber $\beta = 2\pi/0.066$ computed using \texttt{resolvent4py} (black dots). The red crosses denote the eigenvalues in \cite{Cao_2021} for comparison. The eigenvectors associated with~$\lambda_1$, $\lambda_2$ and $\lambda_3$ are shown in figure \ref{fig:ramp_evecs}.
}
\label{fig:ramp_evals}
\end{figure}

\begin{figure}
\centering
\includegraphics[trim=10 10 10 10, clip, width=0.7\linewidth]{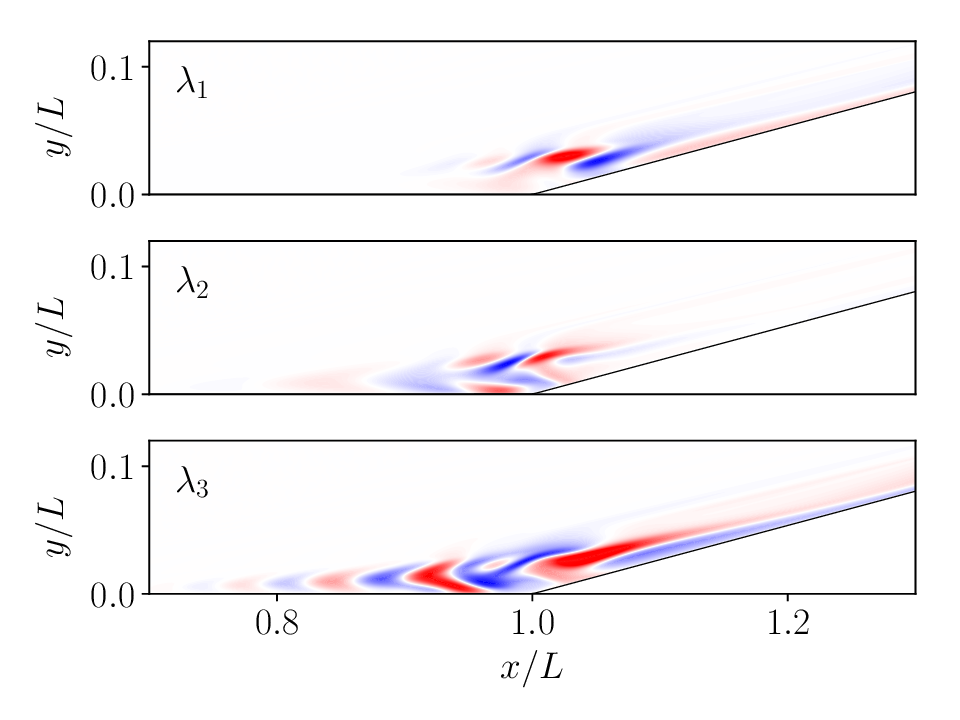}
\caption{Real part of the spanwise velocity component from the eigenvectors corresponding to eigenvalues $\lambda_1$, $\lambda_2$ and $\lambda_3$ in figure \ref{fig:ramp_evals}.}
\label{fig:ramp_evecs}
\end{figure}

The first example is a Mach--$7.7$ flow over a $15\deg$ compression ramp of length $L = 0.1\, m$.
In the freestream, the Mach number is $M_\infty = 7.7$, the pressure is $p_\infty = 760\, Pa$, the temperature is $T_\infty = 125\, K$, the streamwise velocity is $u_\infty = 1726\,m/s$ and the Reynolds number based on $L$ is $Re_{\infty, L} = 4.2\times 10^5$.
This flow configuration is studied extensively in \cite{Cao_2021}, and we refer to the latter for details on boundary conditions, spatial discretization, mesh and non-dimensionalization.
As discussed in \cite{Cao_2021}, this flow admits a two-dimensional steady state solution (see figure \ref{fig:ramp_mach_contours}) that is unstable with respect to three-dimensional perturbations.
These instabilities can be detected by linearizing the governing equations about the aforementioned steady state and studying the dynamics of three-dimensional, spanwise-periodic perturbations with wavenumber~$\beta$.
Upon linearization, the spatially-discretized compressible Navier-Stokes equations can be written compactly as
\begin{equation}
    \frac{d}{dt}q_\beta = A_\beta q_\beta,\quad q_\beta \in \mathbb{C}^N,
\end{equation}
where the state vector $q_\beta$ denotes the wavenumber-$\beta$ Fourier coefficients of the spatially-discretized flow variables.
Here, given a computational grid of size $n_y \times n_x = 236\times 1076$ and $5$ flow variables, the size of the state vector is $N =  5\, n_y\, n_x \approx 1.27\times 10^6$.
The eigenvalues of $A_\beta$ and corresponding eigenvectors for (non-dimensional) wavenumber $\beta = 2\pi/0.066$ are computed using \texttt{resolvent4py} with $240$ processors on Stampede3 at Texas Advanced Computing Center.
The eigenvalues are shown in figure \ref{fig:ramp_evals}, where we see good agreement with the results in Cao et al.~\cite{Cao_2021}. 
Indeed, we find that the flow exhibits multiple instabilities corresponding to the five eigenvalues $\lambda$ with real part greater than zero.
The flow structures associated with these instability modes are shown in figure \ref{fig:ramp_evecs}, where we plot contours of the real part of the spanwise velocity component from the eigenvectors corresponding to three unstable eigenvalues.
Here, we find that the eigenvector corresponding to~$\lambda_1$ exhibits very good qualitative agreement with its counterpart in figure 14a in \cite{Cao_2021}.
The eigenvectors corresponding to $\lambda_2$ and $\lambda_3$ are very similar in nature to those in figures 14b and 14c in \cite{Cao_2021}, but it is important to observe that a direct comparison cannot be made since the modes shown in figures 14b and 14c in \cite{Cao_2021} correspond to wavenumbers $\beta = 0.070$ and $\beta = 0.079$, respectively.

\subsection{Hypersonic flow over a cone}
\label{subsec: cone}

Another illustrative example is a Mach-5.9 flow over a 7 deg half-angle blunt cone. The freestream quantities are $M_\infty = 5.9$, $p_\infty = 3396.3$ $Pa$, $T_\infty = 76.74$ $K$, $u_\infty = 1036.1$ $m/s$, and $Re_{\infty} = 18 \times 10^6$ and the domain length is $L = 0.4893 \, m$. Bluntness effects produce unique instability phenomena in comparison with sharp cones, and are extensively studied using various DNS and operator theoretic methods in receptivity studies ~\cite{nichols_2024}. 
At high Mach numbers, a discrete higher-order mode termed the second Mack mode becomes destabilized and dominates laminar-to-turbulent transition. This mechanism manifests itself as acoustic waves trapped within the boundary layer. 
Likewise, response modes are modulated by the entropy layer observed near the front of the cone, which can contribute to distinct turbulent transition mechanisms. \\

\begin{figure}
\centering
\includegraphics[width=0.75\linewidth]{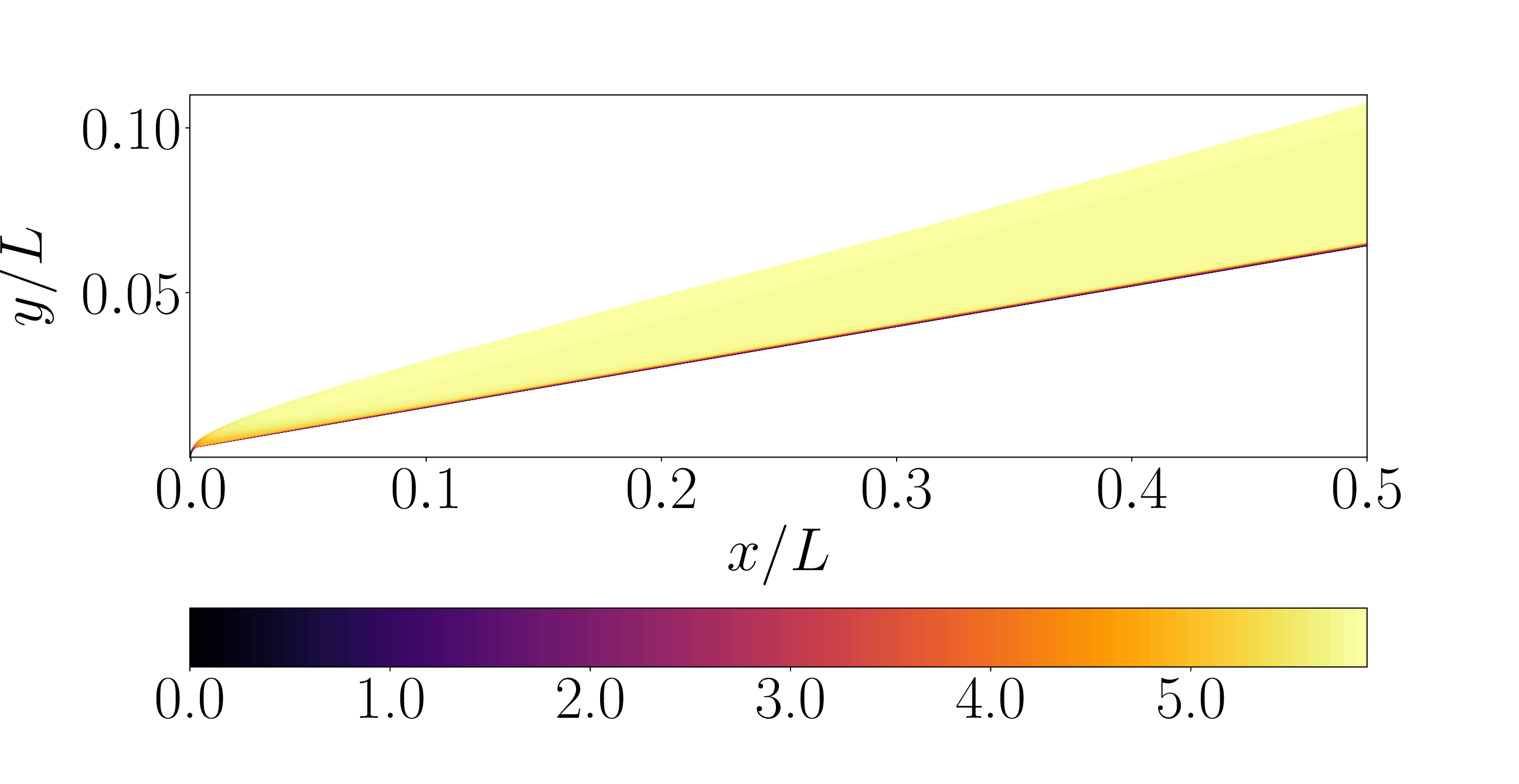}
\caption{Mach number contours of two-dimensional steady state solution for the Mach--$5.9$ flow over a $7\deg$ blunt cone, truncated to solely the post-shock regime and normalized by domain length.}
\label{fig:cone_mach_contours}
\end{figure}

\begin{figure}
    \centering
    \includegraphics[width=0.75\linewidth]{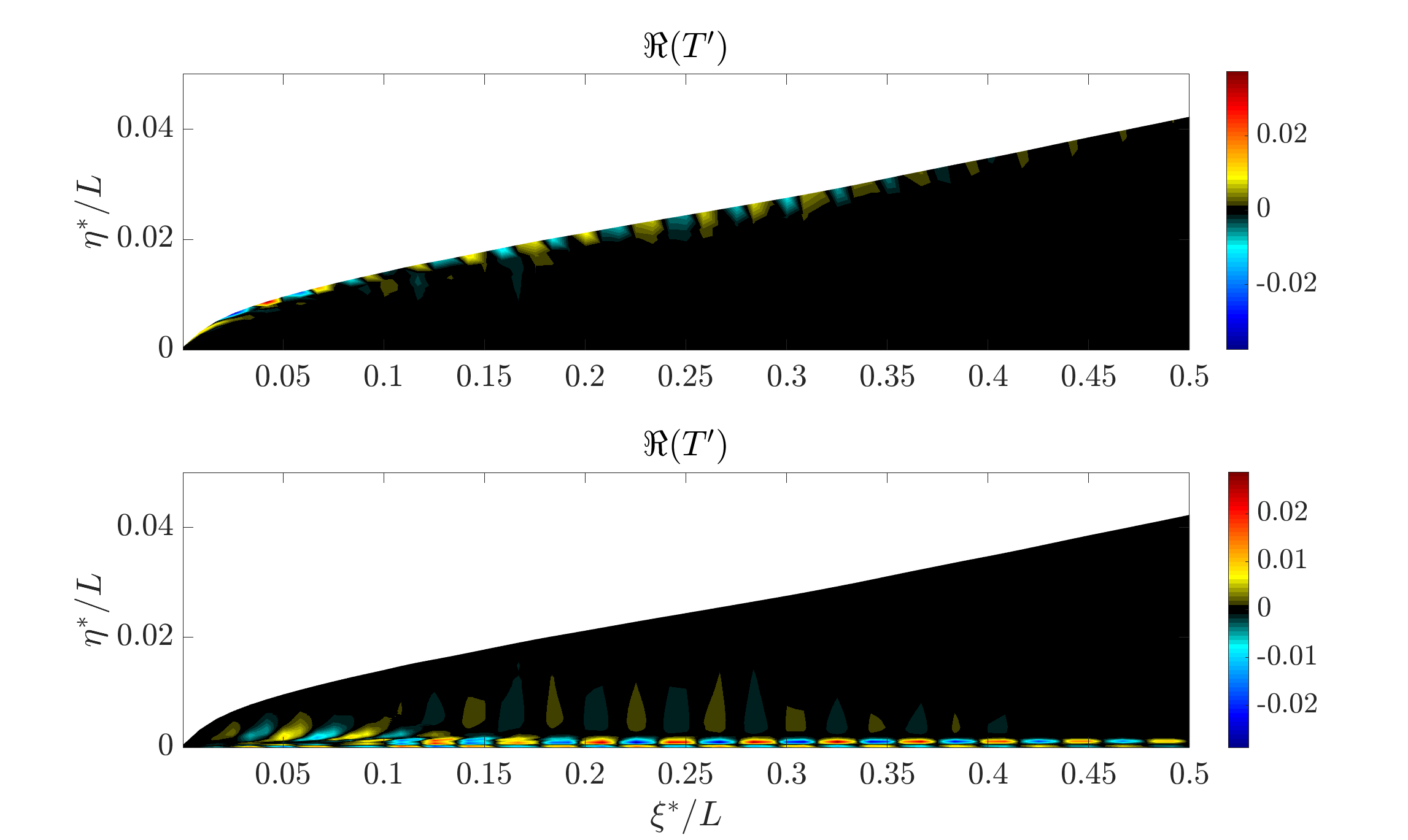}
    \caption{Temperature forcing and response modes computed using RSVD-$\Delta$t plotted in streamwise and wall-normal coordinates normalized by the streamwise length of the domain.}
    \label{fig:cone_temp_forcing}
\end{figure}

The spatially-discretized Navier-Stokes equations are linearized about the base flow shown in figure~\ref{fig:cone_mach_contours}, and the singular value decomposition of the resolvent operator is computed using the RSVD-$\Delta$t algorithm using a script almost identical to the one shown in section \ref{subsec: code_snippet}.
The temperature perturbations extracted from the leading right and left singular vectors of the resolvent are seen in figure ~\ref{fig:cone_temp_forcing}. 
The response mode corresponding to maximal amplification appears trapped in the boundary layer, while the forcing mode follows the decaying entropy layer's contour within the boundary layer, indicative of the importance of the high entropy gradients of the flow due to bluntness.

\section{Impact}

In the fluid mechanics community alone, linear systems tools like stability analysis \cite{Schmid2001stability, Cao_2021, vollmer2025}, resolvent analysis \cite{Jovanovic2005componentwise,McKeon2010critical,Herrmann2021}, open-loop and feedback control design \cite{Semeraro2013riccati,Yeh2019,woo2024}, model reduction of linear time-invariant and time-periodic systems via balanced truncation \cite{Moore1981principal, Rowley2005model, dergham2011, padovan2024}, etc., are featured in hundreds of peer-reviewed publications and industry applications.
However, despite the popularity of these methods and algorithms, there is no unified platform that helps with their streamlined implementation.
\texttt{resolvent4py} aims to bridge this gap by offering a Python package that can significantly reduce the upfront cost associated with developing parallel code to answer pressing research questions in physics and engineering.
Specifically, not only does \texttt{resolvent4py} ship with fully-parallel state-of-the-art algorithms \cite{arnoldi1951, rsvd, ribeiro2020, dergham2011} that are used on a daily basis by researchers and engineers in several branches of physics, but it also offers a user-friendly, pythonic environment that allows for rapid code development, testing and deployment.
In other words, our package can help scientists and engineers solve very large-scale problems of practical interest at a fraction of the human cost that would otherwise be necessary to develop application-specific software.
The simple and self-contained code snippet in section~\ref{subsec: code_snippet} is a testament to the friendly environment enabled by our package.

Additionally, we believe that the current version of \texttt{resolvent4py} can serve as a stepping stone for additional software development---both by the original developers and the scientific community at large---that will be featured in future releases.
For example, the infrastructure of \resforpy enables the straightforward implementation of time-periodic balanced truncation using ``Frequential Gramians'' \cite{padovan2024}, harmonic resolvent analysis via time stepping \cite{farghadan_hr_2024}, wavelet-based resolvent analysis \cite{ballouz2024} and One-Way Navier-Stokes (OWNS) for slowly-developing flows \cite{towne_owns_2022}.
\revone{Furthermore, it should be possible to implement the algorithms discussed in \cite{benner2008, benner2013} for the efficient computation of low-rank solutions to Lyapunov and Riccati equations of large-scale, sparse linear systems such as those that arise from the spatial discretization of partial differential equations.
These solutions could then be leveraged for model reduction and closed-loop control design.
}

\section{Conclusions}

In this manuscript, we introduced \resforpy, a parallel Python package for the analysis, model reduction and control of large-scale linear systems.
Large-scale linear systems with millions or billions of degrees of freedom are ubiquitous in the sciences and engineering, and this package is built to provide a friendly pythonic environment for rapid development and deployment of MPI-based parallel software to solve problems of practical interest. 
Our library leverages \texttt{mpi4py} for distributed-memory parallelism, and it uses the functionalities and data structures provided by \texttt{petsc4py} and \texttt{slepc4py} to enable stability analysis, input/output analysis, model reduction, control and optimization of linear time-invariant and time-periodic systems. 
Additionally, the package ships with several functions and features that are available to the user through the \resforpy namespace and that facilitate user-specific and application-specific software development within the larger \resforpy infrastructure.
\revone{Finally, although the primary focus of this package is on linear systems, we wish to remark that the data structures and functionalities of \resforpy can be extended straightforwardly to accommodate the analysis, model reduction and control of large-scale nonlinear systems.}

\section*{Acknowledgements}

This material is based upon work supported by the National Science Foundation under Grant No. 2139536, issued to the University of Illinois at Urbana-Champaign by the Texas Advanced Computing Center under subaward UTAUS-SUB00000545 with Dr.~Daniel Stanzione as the PI.
DB gratefully acknowledges support from the Office of Naval Research (N00014-21-1-2256), and
CR gratefully acknowledges support from the Air Force Office of Scientific Research (FA9550-19-1-0005).
TC and VA gratefully acknowledge support from The Boeing Company (CT-BA-GTA-1) and the Office of Naval Research (N00014-25-1-2072).
BV was supported by the LDRD Program at Sandia National Laboratories. Sandia is managed and operated by NTESS under DOE NNSA contract DE-NA0003525.
The computations in section \ref{subsec: ramp} were performed on TACC’s Stampede3 under ACCESS allocation CTS090004.



\bibliographystyle{elsarticle-num} 
\bibliography{refs}



\end{document}